\documentclass[12pt]{JHEP3}

 \input{epsf}

 \usepackage{epsfig}

 \title{Geometric construction of elliptic integrable systems and 
${\cal N}=1^{\ast}$ superpotentials}
 \author{S. Prem Kumar \\  Department of Physics \\
University of Wales Swansea \\ E-mail: \email{s.p.kumar@swan.ac.uk} }
\author{Jan Troost \\  Center for Theoretical Physics \\  MIT \\
    77 Mass Ave \\ Cambridge, MA 02139  USA \\E-mail: \email{troost@mit.edu}}
\preprint{ SWAT-326 \\  MIT-CTP-3193 }

 \abstract{We show how the elliptic Calogero-Moser integrable
systems arise from a symplectic quotient construction,
generalising the construction 
 for
$A_{N-1}$ by Gorsky and Nekrasov to other  algebras.
 This clarifies the role of (twisted)
affine Kac-Moody algebras in elliptic Calogero-Moser systems and
allows for a natural geometric construction of Lax operators
for these systems. We elaborate on the connection of the associated
 Hamiltonians to superpotentials for 
${\cal N}=1^*$ deformations of ${\cal N}=4$ supersymmetric
 gauge theory, and argue how non-perturbative physics generates the
elliptic superpotentials.
 We also discuss the relevance of these systems and the
 associated quotient construction 
to open problems in string theory.
In an appendix, we use the theory of orbit algebras to
show the systematics behind the folding procedures for these 
integrable models.} 

\begin{document}

 \setcounter{equation}{0}

\section{Introduction}
Integrable systems are an interesting subclass of 
physical theories to study,  because we can
explicitly calculate a lot of the physical properties of these
theories, and because integrable systems find surprisingly
rich applications in
many other fields, like four-dimensional ${\cal N}=2$ and
${\cal N}=1$
supersymmetric field theory, and string theory.

The specific integrable systems we want to study are
elliptic Calo\-ge\-ro-Moser models. These form the top of
a pyramid 
 of integrable systems (see e.g.
\cite{Olshanetsky:1981dk, Olshanetsky:1983wh}), since the (affine) 
Toda systems, the trigonometric and the rational models can
be derived from them by taking appropriate limits in 
parameter and phase space \footnote{Strictly speaking, the
Calogero Moser systems appear as degenerations of
Ruijsenaars-Schneider systems which can be interpreted in
terms of a five dimensional theory (see 
e.g.\cite{Braden:1999zp} and references therein). }
. Diverse Lax operators with spectral
parameter have been
constructed for these theories using mostly algebraic
methods 
\cite{Krichever, D'Hoker:1998yh, Bordner:1998xs, Bordner:1999xs, Bordner:1999sw},
but a good geometrical understanding of   
the full set of elliptic models is lacking. 

Our primary motivation for studying these models is from the point of
view of deriving superpotentials for ${\cal N}=1$ supersymmetric
gauge theories obtained as mass-deformations of finite ${\cal N}=2$
and ${\cal N}=4$ theories with general gauge groups. The
superpotential for the ${\cal N}=1^*$ theory (the ${\cal N}=1$ preserving
mass-deformation of ${\cal N}=4$) for $SU(N)$ gauge group was shown
to coincide in \cite{Dorey:1999sj} with the elliptic Calogero-Moser
Hamiltonian associated to the $A_{N-1}$ root system. 

A direct
derivation of this proposal was provided in \cite{Dorey:2001qj} via a
hyperkahler quotient construction following the ideas of
\cite{Kapustin:1998xn} wherein the $SU(N)$, ${\cal N}=2^{\ast}$ theory
(the ${\cal N}=2$ preserving mass-deformation of ${\cal N}=4$) was
realised on the worldvolume of a Type IIA brane setup. This
configuration was mapped by a series of dualities onto the Higgs
branch of a ``magnetic'' mirror theory. Following this procedure in
string theory for
other gauge groups is an extremely attractive program, but one which
runs into difficulties due to the lack of knowledge about the
associated brane configurations.
Nevertheless, one can try to gain insight into this approach by
studying the geometry   
of the integrable systems associated to other root systems.
Therefore, following \cite{Gorsky:1994dj},
we want to point out a geometrical
derivation of a certain form of the Lax operators for all elliptic
integrable systems using
a symplectic quotient construction. This also yields an
alternative, geometrical proof of the integrability of the
system. 

In \cite{Gorsky:1994dj} this derivation of the integrable model
and its Lax operator was done for the elliptic Calogero-Moser
model based on the $A_{N-1}$ root system.
In sections \ref{current}, \ref{punctures} and \ref{zerolevel},
we extend their analysis.
Our construction will naturally incorporate affine algebras.
Specifically,  this will help 
in interpreting the role of twisted affine algebras
in twisted elliptic Calogero-Moser models. In section \ref{open}
 we elaborate on the relevance of these integrable
systems for supersymmetric field theory, particularly in connection
with the superpotentials of mass deformed ${\cal N}=4$ theories. We
discuss the various non-perturbative contributions to the superpotentials.
 We
also speculate on the relevance of these models to brane configurations
in string theory, in analogy to the
important role for the (spin) elliptic integrable model corresponding to the
$A_{N-1}^{(1)}$ algebra 
\cite{Dorey:1999sj, Dorey:2001qj, Donagi:1996cf} in this context.
Finally, in appendix 
\ref{foldings} we point out how the
techniques developed in the analysis of fixed point theories in 
conformal field theory \cite{Fuchs:1996zr}
 explain the systematics behind folding procedures in integrable
systems.
\section{Current algebra on the torus}
\label{current}
In this section we describe the symplectic quotient construction
relevant for the elliptic integrable
systems. For the
elliptic Calogero-Moser model based on the root system of the $A_{N-1}$
algebra, this was done in  \cite{Gorsky:1994dj}. We indicate how
to extend that analysis to include all elliptic integrable models.

It is not surprising that we have to start with a current
algebra on the torus, since the complexified elliptic integrable
model has a double periodicity.
Consider then a torus $\Sigma_{\tau}$ with modular parameter $\tau$.
We choose a holomorphic differential $\omega$ such that for the
$\alpha$ and $\beta$ cycle of the torus we have:
\begin{eqnarray}
\int_\alpha \omega &=&1 \\
\int_\beta \omega &=& \tau.
\end{eqnarray}
We will refer to the $\alpha$-direction as direction $x^1$ and
the $\beta$-direction as direction $x^2$ with periodicities
$ \omega_1$ and $ \omega_2$ respectively.

As in \cite{Gorsky:1994dj} we consider the algebra $\bar{g}^{\Sigma_{\tau}}$
of maps $\phi$ from 
an elliptic curve $\Sigma_{\tau}$
 to a complexified simple 
 Lie algebra $\bar{g}$
\cite{Etingof:1994br}.
In order to be able to deal with twisted elliptic integrable
models as well as ordinary elliptic integrable models, 
we need to distinguish two cases. In the first case, the
map $\phi$ is periodic in both directions of the torus:
\begin{eqnarray}
\phi(x^1+\omega_1, x^2)  &=& \phi(x^1,x^2) \nonumber \\
\phi(x^1,x^2+\omega_2)  &=& \phi(x^1,x^2).   
\end{eqnarray}
In the second case, the map $\phi$ is twisted around
 the $\beta$-cycle
of the torus. Suppose $T_a$ forms a basis of $\bar{g}$
and  $\sigma$ is an outer
automorphism of the Lie algebra $\bar{g}$ of order $l$ (namely
$\sigma^l=1$).
Then the map $\phi$ satisfies the 
twisted boundary conditions:
\begin{eqnarray}
\phi(x^1+ \omega_1, x^2) &=& \phi(x^1,x^2)  \nonumber \\
  \phi^a(x^1, x^2+ \omega_2) T_a &=& \phi^a(x^1,x^2) \sigma(T_a).
\end{eqnarray}
A central extension \cite{Etingof:1994br} of the algebra of maps is defined
by our holomorpic differential
 $\omega \in H^{(1,0)}(\Sigma_{\tau})$ and
an $H^{\ast (1,0)}(\Sigma_{\tau})$ valued two-cocycle
\begin{eqnarray}
c(X,Y) &=& \int_{\Sigma_{\tau}} \omega \wedge <X,dY>
\end{eqnarray}
where $X,Y \in \bar{g}^{\Sigma_{\tau}}$ and the brackets $<\, ,>$
 indicate an invariant bilinear form
on $\bar{g}$.
The cotangent bundle of the extended algebra $\hat{\bar{g}}^{\Sigma_{\tau}}$
consists of elements $(\phi, c, \bar{A}, \kappa)$ where
$\phi: \Sigma_{\tau} \rightarrow \bar{g}$, $\bar{A} \in \Omega^{(0,1)}
\otimes \bar{g}$ and $c, \kappa \in \mbox{\boldmath $C$} $. We have a pairing
\begin{eqnarray}
<(\bar{A},\kappa), (\phi,c)> &=& \kappa.c+ \int_{\Sigma_{\tau}}
\omega \wedge \mbox{tr} \phi \bar{A}.
\end{eqnarray}
The current group acts naturally on the cotangent bundle as gauge
transformations, and the action preserves the standard symplectic form on
the cotangent bundle. The moment map for the action on the cotangent
bundle is \cite{Etingof:1994br}:
\begin{eqnarray}
\mu_1 = \kappa \bar{\partial} \phi+ {[} A_{\bar{z}}, \phi {]},
\end{eqnarray}
which naturally takes values in  the dual of the simple Lie algebra
$\bar{g}$ we started out with -- which we identify with
$\bar{g}$ via the Killing form.

\section{Punctures}
\label{punctures}
As in \cite{Gorsky:1994dj, Donagi:1996cf, Hitchin, Markman},
 we can introduce
punctures in the Riemann surface, and extend the moment map to 
include the action of the current group on vector spaces attached
at points of the Riemann surface. We will be interested in
cases where there is only one  puncture 
 in the Riemann surface.

 Before we give the form of the additional
piece in the moment map, we recall a few properties of (twisted) affine
Kac-Moody algebras. In the case of twisted
boundary conditions, the algebra $\bar{g}$ naturally splits into
subalgebras
$\bar{g}_j$ that consist of the eigenvectors of $\sigma$ with eigenvalue 
$e^{i 2 \pi  j/l}$,
as is well-known from the theory of affine Kac-Moody algebras 
(\cite{K, KM, Goddard:1986bp, Fuchs:1992nq})
to which
we refer for more details.\footnote{The affine Kac-Moody algebra
is a subalgebra of the current algebra 
associated to the $\beta$-cycle of the torus. The periodic
functional dependence on $x^1$ is not crucial in the present
section and is implicit
in the following. We will then often denote $x^2$ by $x$.}
Note that the $\bar{g}_j$ are representation spaces for the
$\bar{g}_0$. (Recall that $l$ is the order of the outer
automorphism $\sigma$.)

Thus, we can split the  field $\phi$ according to the periodicity
of its components, and their decomposition in terms of step
and Cartan subalgebra (CSA) generators of $\bar{g}_0$ and weight spaces of the
other $\bar{g}_j$:
\begin{eqnarray}
\phi &=& \sum_{j=0}^{l-1} (\phi_j^r H_r^j + \phi_j^{\lambda} E^j_{\lambda}).
\end{eqnarray}
Here, the index $j$ denotes the periodicity of the component of $\phi^r_j$ 
 in the $\beta$-direction
 while the label $r$ (implicitly summed over)
indexes the  CSA generators or,
alternatively the multiplicity of the zero weight in the $\bar{g}_{0}$
module spanned by $\bar{g}_j$. Similarly, the upper label $\lambda$ 
in the second term denotes roots
or weights of $\bar{g}_0$, depending on whether $j$ is zero or non-zero,
respectively. In the following, we will concentrate on all affine
algebras except $A_{2k}^{(2)}$, for ease of notation only. In that case,
the weight space for $\bar{g}_j$ is always the set of short roots ${\{}
\alpha_s {\}}$.

Next, we motivate an additional piece in the moment map. Firstly, we consider
the following formal
sums over non-CSA generators in the standard realisation
of the current algebra.
 For an 
untwisted current algebra, we sum \footnote{Notation: $\Delta$ for
the root system of $\bar{g}_0$, $\alpha$ for roots 
and $\alpha_{l,s}$ for the long and
the short roots of $\bar{g}_0$.}:
\begin{eqnarray}
\sum_{\alpha \in \Delta,n \in \mbox{\boldmath $Z$}}
 E_{\alpha}^0 \otimes e^{inx} &=& \delta_1(x)
 \sum_{\alpha} E_{\alpha}^0
\end{eqnarray}
where $\delta_l(x)$ is a $\delta$-function with periodicity $2\pi l$.
For all twisted algebras (except for $A^{(2)}_{2k}$) we consider the sums:
\begin{eqnarray}
\sum_{\alpha_l,n} E^0_{\alpha_l} \otimes e^{inx} &=& \delta_1(x)
 \sum_{\alpha_l} E_{\alpha_l}^0 \label{twistedRHS} \\
\sum_{\alpha_s,n} E^j_{\alpha_s} \otimes e^{i(ln+j)x/l}
&=&
   (\delta_l(x)+\dots+e^{2 \pi i (l-1)j/l} \delta_l(x+(l-1))
\nonumber \\
&&
\sum_{\alpha_s} E^j_{\alpha_s}, \nonumber
\end{eqnarray}
where $j$ now runs from $0$ to $l-1$.
For the algebra  $A^{(2)}_{2k}$ we will not write out the detailed
formulas, but  a similar computation would naturally involve the
three Weyl orbits in the affine root system.

Now, we recall that to recover the trigonometric integrable model in
\cite{Gorsky:1994pe},
the extra piece in the moment map ($\mu_2$ or RHS) was chosen such that
all non-CSA generators were weighted equally. We will take a similar
 road in this current algebra system, and our choice will
turn out to be the appropriate one to recover the standard
elliptic integrable models.\footnote{Note that
is also equivalent to the choice of moment map made in \cite{Donagi:1996cf}
for the $A_{N-1}$ elliptic integrable model.}
Thus, 
when we consider a sum over roots as our RHS (and sum over the KK
modes in the $x^1$-direction in a similar manner) for the untwisted case,
we choose to weigh all contributions 
equally with weights $\nu$, and find a RHS:
\begin{eqnarray}
\mu_2 &=& - \nu \sum_{\alpha \in \Delta} E^0_{\alpha} \delta_1(z).
\end{eqnarray}
For the twisted case, we weigh the contributions from each Weyl orbit
of $\bar{g}_0$ given in (\ref{twistedRHS}) with weights $\nu_{l,s}$.
Strictly speaking, it is necessary to
work out in detail how these choices for the moment
map
arise from a
symplectic quotient of a vector space inserted at the 
puncture(s).\footnote{In the light of the later
remarks on realizing these models using brane configurations, 
a careful analysis
of these vector spaces should reveal information on the open string degrees
of freedom living on D-branes in the presence of $ON$-planes.}
Our educated guess will turn out to yield the right results.
It would be interesting
to include more punctures and extend this analysis to Gaudin and spin
models (see e.g. \cite{Nekrasov:1996nq} \cite{Inozemtsev:2001ux} ).
\section{The zero level submanifold}
\label{zerolevel}
We want to study the zero level submanifold $\mu_1+\mu_2=0$. We will
treat the twisted case (except for $A^{(2)}_{2k}$) 
in detail -- the first, untwisted
case has an easier analogue.
First of all, 
using the current group, we assume we can   conjugate 
 $A_{\bar{z}}$ to a constant element of the CSA of 
$\bar{g}_0$.\footnote{This is always possible for untwisted 
affine Kac-Moody algebras as proven in \cite{Frenkel}. For twisted affine
algebras there is no similar theorem known to us.}
The equation for the 
level zero submanifold splits nicely into different parts. The equations 
 for the generators with zero weight under  $\bar{g}_0$, are:
\begin{eqnarray}
\kappa \bar{\partial} \phi^r_0 &=& 0 \nonumber \\ 
\kappa \bar{\partial} \phi^s_j &=& 0.  \quad (j\neq0)
\end{eqnarray}
The solutions are $\phi^r_0 = p^r=\mbox{constant}$ for the first
equation and $\phi^s_j = 0$ for the second, since that field component
has to be both constant and satisfy twisted boundary conditions.
 Note that our choice for
the total moment map was judicious, in that a single puncture on the
CSA would have lead to a contradictory equation for the periodic
excitations.\footnote{The abscence of CSA components in the
moment map may be reminiscent of the 
nature of the algebraic approach in \cite{Bordner:1999us}, based on
representations of the Coxeter group (generated by the roots only).}

For the other weights the equations read:
\begin{eqnarray}
\kappa \bar{\partial} \phi^{\alpha_l}_0
+ <a,\alpha_l>  \phi^{\alpha_l}_0 &=& 
\nu_l \delta_1 (z) \label{Higgsbranchstep}  \\
\kappa \bar{\partial} \phi^{\alpha_s}_j
+ <a,\alpha_s>  \phi^{\alpha_s}_j &=& 
\nu_s  (\delta_l(z)+\dots+e^{2 \pi i (l-1)j/l} \delta_l(z+(l-1)i))
\nonumber
\end{eqnarray}
Here we chose $A_{\bar{z}}= a \in \bar{h}_0$, the CSA of $\bar{g}_0$, and
for simplicity we put  $ \omega_1=1$ and
$\omega_2=i$.
To solve these equations, we introduce new variables $\psi$ which
are related to the $\phi$ components as:
\begin{eqnarray}
\phi^{ \alpha_l}_0 &\equiv& \exp( \pi  <a,\alpha_l> (z- \bar{z})/ \kappa \tau_2)
\psi^{\alpha_l}_0 \nonumber \\
\phi^{ \alpha_s}_j &\equiv& \exp( \pi  <a,\alpha_s> (z- \bar{z})/ \kappa \tau_2)
\psi^{\alpha_s}_j. 
\end{eqnarray}
These new fields are meromorphic functions on the torus, and they satisfy
boundary conditions:
\begin{eqnarray}
\psi^{\lambda}_{j} (z+1) &=&
 \psi^{\lambda}_{j} (z) \\
\psi^{\lambda}_{j} (z+ i) &=& e^{ 2 \pi i j/l}
e^{- \frac{2 \pi i}{\kappa} 
 <a,\lambda>} \psi^{\lambda}_{j} (z). \quad \mbox{(any $j, \lambda$)}
\end{eqnarray}
For the long root excitations periodic in the $x^2$ direction,
 the solution to the boundary conditions and
pole structure are given by \footnote{See appendix  \ref{theta} for our
standard $\theta$-function conventions.}:
\begin{eqnarray}
\psi^{\alpha_l}_0 &=& \nu_l' 
\frac{\theta_{11}(z+ <a,\alpha_l>/\kappa)}{\theta_{11}(z)
\theta_{11} ( <a,\alpha_l>/ \kappa)},
\end{eqnarray}
where $\nu'_l$ is related to $\nu_l$ via the residue of the
$\theta_{11}$ function at $0$.
The solution for the short root excitations reads 
\begin{eqnarray}
\psi^{\alpha_s}_j &=& \nu_s'
\frac{\theta_{11}(z+ \frac{j \omega_2}{l}+ <a,\alpha_s>/\kappa)}{\theta_{11}(z)
\theta_{11} ( \frac{j \omega_2}{l}+ <a,\alpha_s>/ \kappa)}.
\end{eqnarray}
The Lax operator of the elliptic integrable system can then
be identified with $\phi$, and the quadratic
Hamiltonian $\mbox{Tr} (\phi^2)$ of the system
is then given by (up to an unimportant constant):
\begin{eqnarray}
H_2 = \frac{p^2}{2} &+& \nu_l^2  \sum_{\alpha_l}{ \cal P}( <a,\alpha_l>)
\nonumber \\ &+ & 
  l \, \nu_s^2 \sum_{\alpha_s} \sum_{j=0}^{l-1}
 { \cal P}( <a,\alpha_s>+  \frac{j \omega_2}{l}  )\label{ham}
\end{eqnarray}
which matches with
the (twisted) elliptic  Hamiltonians of \cite{Bordner:1999xs}.
Note that the derivation was made with a twist in the $\omega_2$
direction (because of standard conventions on $\theta$-functions). 
We might as well have taken the twist to be in any other
cycle of the torus (and in particular the cycle associated to $\omega_1$).
 As we already remarked, it would be interesting
to generalize  the geometric derivation of Lax
operators and Hamiltonians to other integrable systems by generalising
our choice of moment map.

\section{Superpotentials for ${\cal N}=1^{\ast}$ theories}
\label{open}
One of our primary motivations for studying the geometry of the elliptic
integrable systems is based on the intimate relationship between classical 
integrable models in two dimensions and the Coulomb branch of
supersymmetric gauge theories, 
noted first in \cite{Gorsky:1995zq, Martinec:1995by}. This
relationship was made precise in  
the work of \cite{Donagi:1996cf} where it was argued
that the spectral curve of an integrable model, namely the $SU(N)$
Hitchin system 
coincides with the Seiberg-Witten curve for
$SU(N)$, ${\cal N}=2$ SUSY gauge theory with a massive adjoint
hypermultiplet 
(known as the ${\cal N}=2^{\ast}$ theory {\it i.e.} an ${\cal N}=2$
preserving mass 
deformation of ${\cal N}=4$ theory). In particular, the moduli of the
Donagi-Witten curve which are the gauge-invariant order parameters on
the Coulomb branch of the ${\cal N}=2^{\ast}$ gauge theory are
identified with the  
Hamiltonians of the integrable model. The
$SU(N)$ Hitchin system was also identified with the elliptic
Calogero-Moser model 
associated to the $A_{N-1}$ root system in
\cite{Martinec:1995qn}. This connection between elliptic
Calogero-Moser systems and ${\cal N}=2^{\ast}$ theories has been
extended to arbitrary gauge groups ${\cal
G}$ in \cite{D'Hoker:1998yi,Itoyama:1995nv}.  

\subsection{${\cal N}=1^{\ast}$ superpotential - from field theory}
An important consequence of the above connection between the 
Hamiltonians of the Calogero-Moser system and the ${\cal N}=2^{\ast}$,
$SU(N)$ gauge theory is that the vacuum value of the
superpotential for the corresponding 
${\cal N}=1^{\ast}$ theory (${\cal N}=1$ preserving mass-deformation
of ${\cal N}=4$ theory) then coincides with the
quadratic Hamiltonian of 
the integrable model \cite{Dorey:1999sj}. Several direct checks of
this superpotential were given in \cite{Dorey:1999sj}. In particular
it was shown that a class of equilibrium configurations of the superpotential
were in one to one correspondence with the massive vacua of $SU(N)$,
${\cal N}=1^{\ast}$ gauge theory. 

It is natural to guess that the above conclusion extends to
superpotentials for ${\cal N}=1^*$ theories with other gauge groups as
well. Specifically, the superpotentials would correspond to the
quadratic Hamiltonians 
of elliptic Calogero-Moser models 
associated to the corresponding root systems \cite{Dorey:1999sj, D'Hoker:1999ft,
Hanany:2001iy}. The general argument for this follows from
viewing ${\cal N}=1^{\ast}$ gauge theory with gauge group ${\cal G}$
as softly-broken ${\cal N}=2^{\ast}$ gauge theory with the same gauge
group and hypermultiplet mass $\nu$. 
The ${\cal N}=2^{\ast}$ theory
has a Coulomb branch where the 
effective superpotential vanishes. On soft breaking to ${\cal
N}=1^{\ast}$ via a mass term for the adjoint chiral multiplet $\Phi$
in the
${\cal N}=2$ vector multiplet, the theory acquires a superpotential,
which for small $\mu$ has the form
\begin{equation}
W_{eff}=\mu \langle{\rm Tr}\Phi^2\rangle =\mu\;u_2,
\end{equation}
where we define $u_k\equiv\langle {\rm Tr}\Phi^k\rangle $ as the
gauge-invariant order-parameters on the Coulomb branch. As in
\cite{Seiberg:1994rs}, the above superpotential can be argued to be
exact, {\it i.e.} valid for large $\mu$ as well. The ${\cal
N}=2^{\ast}$ gauge theory has a $U(1)_J$ global symmetry which is a
subgroup of the $SU(2)_R$ symmetry. The scalar components of  
the ${\cal N}=2$ hypermultiplet carry charges $+1$ 
under $U(1)_J$, while the ajoint scalar $\Phi$ is neutral under this
transformation. The ${\cal N}=1^*$ theory inherits this $R$-symmetry
provided the mass parameter $\mu$ is assigned a charge $+2$. Since the
superpotential must have $R$-charge $+2$, the only term with this charge and 
consistent with the requirement of holomorphy and analyticity 
in the variables, $u_k, \mu$ and $\nu$ is $\mu u_2$ which must
therefore be the exact value of the low-energy superpotential. The
correspondence between elliptic Calogero-Moser models and ${\cal N}=2^\ast$ gauge
theories identifies the gauge-invariant order parameters $\{u_k\}$
with the conserved Hamiltonians (action variables) of the associated
integrable models. In particular, $u_2$ is directly identified with
the quadratic Hamiltonian of the integrable system. 

Thus the  superpotential for the ${\cal
N}=1^*$ theory with gauge group ${\cal G}$ is 
\begin{eqnarray}
W_{eff}&=&\mu \;u_2 \label{wsuperpot}\\\nonumber
&=&\mu (\nu_l^2\sum_{\alpha_l}{ \cal P}( <a,\alpha_l>)
 + l\nu_s^2\sum_{j=0}^{l-1} 
\sum_{\alpha_s}{ \cal P}( <a,\alpha_s>+  \frac{j \omega_1}{l} ) )
\end{eqnarray}
where we have simply replaced $u_2$ with the Hamiltonian of the
elliptic integrable model associated to the algebra of ${\cal G}$,
derived in Eq.(\ref{ham}). Note that we have used the twisted
Hamiltonians where the twist has been performed in the $\omega_1$
direction. This choice was made in anticipation of the periodicity
properties of the physical degrees of freedom from the gauge theory
viewpoint. 
The complexified coupling constant of the underlying ${\cal N}=4$
theory coincides with the complex structure $\tau$ of the torus
$\Sigma_\tau$ on which the Weierstrass functions are defined.
It is important to note that the twisted Calogero-Moser systems
naturally make an appearance for non-simply laced ${\cal G}$
\cite{D'Hoker:1998yh}. We further remark that the weighting factors
$\nu_{l,s}$ are fixed in terms of the field theory parameters to be 
$\nu_l=\nu$, the hypermultiplet mass and $\nu_s=\nu/l$. These
parameters can be fixed by comparing the conjectured elliptic
superpotential in a certain limit (the affine Toda limit to be
discussed below) to explicit field theory computations of the
superpotential in that limit.

The dynamical variables $a$, of the integrable system  have a natural physical,
gauge theory interpretation when the ${\cal N}=1^*$ theory is compactified on $R^3\times
S^1$. In the latter context $a$ represent a complex combination of the
Wilson lines and the dual photons of the 3D effective theory
\cite{Dorey:1999sj}. In particular, in the classical ${\cal N}=1^*$
theory on $R^3\times S^1$, the Wilson line $a_1$ around the $S^1$ is a
modulus and may be chosen to lie in the Cartan subalgebra. Generic
VEVs for the Wilson lines break ${\cal G}\rightarrow U(1)^r$ where
$r={\rm rank}\;({\cal G})$. Symmetry of the theory under large gauge 
transformations ({\it i.e.} gauge transformations that twist around the
$S^1$ by an element of the center of ${\cal G}$), requires the Wilson
lines to be periodic variables 
under $a_1 \rightarrow a_1+2\pi {\bf \omega}^*$. Here ${\bf \omega}^*$ is
an element of the co-weight lattice. In
addition, in the 3D effective theory we may exchange the $r$ photons
for dual scalar fields $a_2$ which are also periodic due to the quantization
of magnetic charge, under $a_2 \rightarrow a_2 + 2\pi {\bf
\omega}$ with $\bf {\omega}$ an element of the weight lattice. 
Supersymmetry then allows us to combine these
scalars into a complex scalar field $a=i(\tau a_1+a_2)$ which forms the
lowest component of a corresponding chiral superfield. The
low-energy, chiral sector of the theory {\it i.e.} the superpotential
is then expected to be a holomorphic function of this periodic
variable.

As we explain
below the elliptic superpotential in Eq.(\ref{wsuperpot}) arises from
semiclassical configurations corresponding to  
3D-monopoles, carrying topological charge in general, with
action proportional to ${\rm exp}(<a,\alpha^*>)$, indicating that the
effective superpotential is a holomorphic function of
$<a,\alpha^*>$. Importantly, 
when $\alpha$ is a long root, a straightforward consequence of
the periodicities of $a_1$ and $a_2$ is that $<a,\alpha_l^*>=<a,\alpha_l>$ must
be periodic on the torus with complex structure $\tau$. In addition,
on general grounds $u_2$ must have modular weight 2
\cite{Donagi:1996cf}. Thus terms 
proportional to ${\cal P}(<a,\alpha_l>)$ must naturally appear in the
superpotential as a consequence of ellipticity on $\Sigma_\tau$. On the other
hand, for short roots $\alpha_s$, the periodicity of the variable
$<a,\alpha_s>$ is different. In particular $<a,\alpha_s>$ is a periodic variable with
periods $\omega_2$ and $\omega_1/l$ where $l=2/\alpha_s^2$. This
explains the appearance of the twisted Weierstrass functions which
involve a sum over the standard Weierstrass functions with arguments 
shifted by multiples of $\omega_1/l$  as in Eq.(\ref {wsuperpot}) leading to the required
periodicity. The twisted functions also ensure that in the
semiclassical limit \cite{D'Hoker:1999ft}, they give rise to terms
proportional to ${\rm exp}<a,\alpha_s^*>$.  Finally, as 
$u_2$ is a dimension two operator in the ${\cal N}=2^*$ theory where
the only mass scale is $\nu$, $u_2\propto \nu^2$. This explains the
mass dependence of the ${\cal N}=1^*$ superpotential $W_{eff}=\mu u_2$.

For generic root systems, it turns out to be a difficult task to
gather direct evidence for the superpotential. In particular, the
classification of the extrema of this superpotential is 
difficult in general. Such a classification and subsequent comparison
with semiclassical predictions (obtained by combining the 
classical analysis of vacua in \cite{Naculich:2001us} with the
associated Witten indices), would provide a
strong test of the elliptic superpotential. 
In the $A_{N-1}$ case, the classification of
massive ${\cal N}=1^{\ast}$ vacua is particularly simple and elegant. Each
such vacuum simply corresponds to an extremum of $W_{eff}$ wherein the
$N$ $a$'s form a lattice $\Gamma^\prime$ on the torus
$\Sigma_\tau$. \footnote{Specifically, if the torus $\Sigma_\tau$ is defined by
the lattice $\Gamma$, then $\Gamma$ must be an order $N$ sublattice of $\Gamma^\prime$}
The total number
of such configurations $=\sum$ divisors of $N$, coincides with the
semiclassical vacuum counting.

\subsection{Trigonometric limit}
Some evidence in
favor of this superpotential for generic ${\cal G}$ may be obtained by
examining the trigonometric limit. This is the limit in which the 
adjoint hypermultiplet with mass $\nu$ is decoupled, keeping fixed the effective
dynamical scale of the 4D 
${\cal N}=2^{\ast}$ 
theory {\it i.e.} $\nu\rightarrow\infty$ and $\tau\rightarrow i\infty$ with 
$\Lambda^2 =\nu^2{\rm exp}(2\pi i \tau/c_2({\cal G}))$ fixed. Here
$c_2({\cal G})$ is the dual Coxeter number for gauge group ${\cal
G}$. In this limit the 4D theory reduces to softly
broken pure ${\cal 
N}=2$ SUSY 
Yang-Mills, and the (twisted) elliptic superpotential above reduces precisely
to the superpotential found in \cite{tim} for ${\cal N}=1$ SUSY Yang-Mills on
$R^3\times S^1$, provided we choose the parameters $\nu_l=\nu$ and $\nu_s=\nu/l$. As discussed in \cite{tim} the latter superpotential
assumes the usual form of the affine Toda potential after a
rescaling and redefinition of variables. As shown in detail in
\cite{tim}, this affine Toda superpotential arises from the $r+1$
types of fundamental BPS monopole contributions in the 3D effective theory on $R^3\times
S^1$. $r$ of these monopoles carry charges $\alpha^*$ in the co-root lattice
where $\alpha$ is a simple root and have action $\sim {\rm
exp}(<\alpha^*,a>)$. In addition there is one type of fundamental
monopole that carries negative magnetic charge given by the lowest
root $\alpha_0^*$ and one unit of 4D
-instanton charge with action $\sim {\rm exp}(<\alpha_0^*,a>+2\pi
i\tau)$. 

\subsection{Semiclassical configurations}

At weak-coupling, the superpotential of the ${\cal N}=1^\ast$ theory
on $R^3\times S^1$ must arise from semiclassical, non-perturbative
contributions. 
As usual, in the 3D effective theory we expect to have BPS monopoles
carrying charges in the co-root lattice 
with action
$\sim {\rm exp}(k<\alpha^*,a>)$ associated to some positive root
$\alpha$. In the four supercharge, ${\cal N} =1^*$ theory, each of
these BPS monopoles has only two supersymmetric fermion zero
modes. All additional zero modes whose existence is predicted by the Callias
index theorem must be lifted as they are not protected by
supersymmetry. Such a lifting of zero modes that are not protected by
SUSY was demonstrated explicitly  
for the sixteen supercharge theory in 3 dimensions in
\cite{nicklift,frasertong}. Hence all these monopoles with two exact zero modes
will contribute to the low-energy ${\cal N}=1^*$ superpotentials. Note
that since the monopole charge can be any positive root, including
non-simple roots they include configurations that are charged under
two different magnetic $U(1)$'s. This situation is in contrast to the
affine Toda superpotential of \cite{tim} where only monopoles
corresponding to simple roots could contribute to the superpotential.

The theory on $R^3\times S^1$, also has BPS
semiclassical configurations carrying both  
magnetic charge and 4D-instanton or topological charge. These
configurations are obtained by shifting the asymptotic value of the
Wilson line variable
$<a_1,\alpha^*>$ by a multiple of its period i.e. $2\pi n$. Such 
configurations contribute terms proportional to 
${\rm exp}(k<\alpha^*,a>) {\rm exp}(2\pi i kn\tau)$. The presence of such
topologically charged BPS states also permits the existence of
well-defined, semiclassical solutions with magnetic charge given by a
{\it negative} 
root and {\it non-zero} 4D topological charge, with action
proportional to 
${\rm \exp}{(-k<a,\alpha^*>)}{\rm\exp}{(2\pi i kn\tau)}$. It must
be emphasized that these are {\it not} anti-monopoles. 
The appearance of such states with negative magnetic charge in the
theory on $R^3\times S^1$ has an elegant description  in the D-brane
picture of the sixteen supercharge theory \cite{janami}. In this
picture, fundamental 
monopoles (associated to simple roots) correspond to 
D1-branes stretching between neighboring D3-branes in a stack of $N$
parallel, separated D3-branes. However, when the D3-branes are placed
on a transverse circle, a new kind of fundamental monopole appears,
stretching between the first and the last D3-brane, associated to the
lowest root of the corresponding affine algebra. Instanton charge is
associated to D1-branes that wind all around the compact direction and
can be related by T-duality to D0-branes dissolved in D4's
i.e. instantons of the corresponding theory. Various
combinations of such D1-brane segments can give rise to all
possible monopole configurations discussed above.

Finally, there
are of course contributions from ordinary 4D instantons as well.
Once again, all 
these configurations carrying 4D instanton number have only two exact
fermion zero modes which are protected by SUSY and are expected to
contribute to the superpotential. Therefore, based on these general
arguments, we expect the low-energy ${\cal N}=1^*$ superpotential to
be an elliptic function with the following semiclassical expansion,
\begin{eqnarray}
W_{eff}=&& \nu^2\mu\sum_{n=1}^{\infty}a_n e^{2\pi i
n\tau}+\nu^2\mu\sum_{\alpha}\left[\sum_{k=1}^{\infty}b_{k,\alpha}
e^{k<a,\alpha^*>}+\right.\\\nonumber 
&&\left.+\sum_{k=1}^\infty\sum_{n=1}^\infty e^{2\pi i k n \tau}
(c_{k,n,\alpha}e^{k<a,\alpha^*>}+d_{k,n,\alpha}e^{-k<a,\alpha^*>})\right]. 
\end{eqnarray}
The semicalssical expansion of the (twisted) elliptic superpotentials
in Eq.(\ref{wsuperpot}) has precisely this form as can be easily seen
by expanding out the Weierstrass functions. 
\subsection{The superpotential from branes and a quotient construction}

Perhaps the most direct way of obtaining the elliptic superpotential for
${\cal N}=1^*$ theory is via the use of mirror symmetry on the field
theory realised on the world-volume of a Type IIA brane setup. This
involves a four-dimensional version of mirror symmetry introduced in
\cite{Kapustin:1998xn, Kapustin:1998pb} and was used to derive
${\cal N}=1^*$ superpotentials for the
$A_{k-1}$ quiver models in \cite{Dorey:2001qj} which include the
$SU(N)$, ${\cal N}=1^*$ theory. The key 
ingredient was the brane configuration \cite{Hanany:1997ie,
Witten:1997sc} of intersecting D4-branes and NS5-branes compactified 
in the conventional $x^6$ and $x^3$ directions. Using Type IIA/IIB
dualities, the Coulomb
branch of the ``electric'' mass-deformed $SU(N)$
theory was mapped to the Higgs branch of a
an $SU(N)$ gauge theory on $R^3\times T^2$ with impurities (punctures
with attached vector spaces)
on the torus $T^2$. This torus has complex structure $\tau$ and must
be identified with $\Sigma_\tau$ introduced earlier.
The Higgs branch equations of the mirror theory are given in terms of
the gauge field on 
the torus $A_{\bar z}$ and the adjoint scalar $\phi$, by D-flatness
equations which  
are precisely Hitchin's equations reduced to 2 dimensions. (The Wilson
lines of the gauge field $A_{\bar z}$ around the two cycles of
$\Sigma_\tau$ are precisely the dual photons and Wilson lines of the
3D effective theory on $R^3\times T^2$). As
discussed in \cite{Dorey:2001qj, Kapustin:1998xn} these conditions can
be thought of as moment map equations for a hyperkahler quotient. In
particular the moduli space of the Higgs branch equations is precisely
a hyperkahler quotient with respect to 
the group of smooth maps $\Sigma_\tau\rightarrow U(N)$ at the zero level of the
moment map. It was shown that this 
moduli space 
corresponds to the zero level manifold associated to the $A_{N-1}$ elliptic
Calogero-Moser system. The geometric construction of elliptic
integrable models is thus very naturally realised in the context of
brane setups in string theory. The symplectic quotient
construction presented in this paper
generalises the above to
integrable models associated to general root systems.

The problems in trying to extend the 
$N=2^{\ast}$ configuration involving spiraling
\cite{Witten:1997sc} D4-branes to other
simple gauge groups are well-known \cite{Uranga:1998uj}\cite{Ennes:1999fb}.
In \cite{Uranga:1998uj} brane configurations realizing
the $N=2^{\ast}$ theory involving O6-planes and D4- and NS5-branes
were put forward, but their corresponding curves 
\cite{D'Hoker:1998yi}
proved difficult to
read off from the brane geometry. In the spirit of
 \cite{Kapustin:1998pb} one could try to find a corresponding mirror
theory, then to study the Higgs branch of the impurity theory on the
torus. This program is attractive, but  runs into difficulties
as the dual (IIB) theory will involve $ON5$-planes (see e.g. 
\cite{Hanany:1999sj} for properties of the $ON5^0$-plane), 
S-dual to $O5$-planes,
and the properties of these planes are in general insufficiently studied to
determine the exact mirror impurity theory.\footnote{We thank
Ami Hanany for interesting discussions on this topic.} 
It would be interesting
to study these theories further, from string theory, as well as using
inspiration from the geometry of the integrable systems worked
out in this paper. In particular,
one could guess that the mirror impurity theory would have a zero level
manifold corresponding to an integrable system treated in this letter,
although this remains to be demonstrated. 

\section{Conclusions and future directions}
\label{conclusions}
A lot can be said on the open problem posed in Section 5.4, but we restrict ourselves to a few more remarks. One can try to move forward on
 the integrable system side of the problem, using a 
correspondence between folding procedures and orientifold planes
\cite{Gorsky:1999gx} (and in particular, by sharpening the dictionary between
the algebraic and geometrical quantities associated to orientifold planes),
 and the analysis of T-duality and S-duality in
the theory of integrable systems (e.g. \cite{Fock:2000ae}). Another issue
that would come into play would be
the
suitability of the form of the Lax operators for taking a
pure $N=2$ limit \cite{D'Hoker:1998yh}\cite{Khastgir:1999pd}. 
As mentioned earlier, on the field theory side,
one needs to check whether the semi-classical analysis of the number of
vacua of the $N=1^{\ast}$ theory  \cite{Naculich:2001us}\footnote{We
thank the authors of this paper for explaining their results.}
 agrees with a possible guess for the
quantum superpotential as was done for the $SU(N)$ gauge group in
\cite{Dorey:1999sj}. 
This involves determining the classical minima
of the potential of these elliptic integrable system which seems to 
be an unsolved and non-trivial problem. It should yield an
interesting picture for the phases of $SO(N)$ and $Sp(2N)$ mass-deformed
gauge theories, and their transformation properties under the duality group.
As described in the previous section, in the $SU(N)$ case,
classification of certain minima corresponding 
to massive vacua of the ${\cal N}=1^\ast$ theory requires classifying
certain lattices. This picture is also directly reflected in the
Donagi-Witten curves for the ${\cal N}=2^*$ theory wherein at certain
points on the moduli space which directly descend to massive ${\cal
N}=1^*$ vacua upon perturbation, the curve degenerates into an unramified $N$
-fold cover of the torus $\Sigma_\tau$. Thus the massive vacua can be
located by finding all the possible genus one $N$-fold covers of
$E_\tau$, a problem that is 
identical to the classification of lattices $\Gamma^\prime$ such that 
$\Gamma$ (the lattice defining the torus $\Sigma_\tau$) is an order
$N$ sublattice of $\Gamma^\prime$. It would be 
extremely interesting to understand if and how this picture
generalises to other gauge groups. Note also
that one can read off the stable solutions for the integrable system
(i.e. the minima for the potential)
directly from the explicit solution for the $A_{N-1}^{(1)}$ elliptic model
\cite{Krichever}, confirming the above picture. 
Unfortunately, no analogous solution for the other
elliptic Calogero-Moser models is known.

Furthermore, by analogy to the $SU(N)$ case \cite{Dorey:2001qj},
 one can speculate on the
relevance of other spin systems (e.g. \cite{Inozemtsev:2001ux}) 
associated to other root systems to
field theories with product gauge groups, but in the light of the difficulties
sketched above it seems to early to make this analogy stick. We 
point it out to show that there remains much ground to be covered.

In summary, in this paper we have analysed the geometry of elliptic integrable
systems following \cite{Gorsky:1994dj}. We clarified the role of
(twisted) affine subalgebras in these integrable systems.
We moreover laid bare some of the systematics of the folding procedures
used to obtain new integrable systems from old ones, using the systematic
results derived in the context of fixed point conformal field theories.
We pointed out possible extensions of our work, which include the 
geometric interpretation to  Gaudin and spin models, and we elaborated 
on the relevance of our analysis to supersymmetric field theories
and on the possible relevance to brane configurations in string theory. We believe we pointed out
interesting directions for future research in this context.
In general, we again demonstrated the cross-fertilisation of current algebra, 
integrable system physics, conformal field theory, 
supersymmetric field theory and string theory and tried to knit them 
tighter together.

\acknowledgments
We would like to thank Nick Dorey, Ami Hanany, Tim Hollowood, Steve Naculich,
David Olive, Howard Schnitzer and Niclas Wyllard for useful discussions
and correspondence. This work is supported in part by funds provided by the
U.S. Department of Energy (D.O.E.) under cooperative research
agreement DE-FC02-94ER40818.

\appendix
\section{Foldings and orbit algebras}
\label{foldings}
This technical appendix makes the link between
the folding procedure of  \cite{Bordner:1999xs}
and the framework of orbit algebras of \cite{Fuchs:1996zr}.
This appendix is separate from the main line of development of the paper,
but we believe it is useful because it sheds more light on
 the systematics behind the folding procedures which were put to good use in 
integrable models, and specifically in elliptic Calogero-Moser models. 
We hope it will also turn out to be useful to clarify the relation 
between the geometric properties of orientifold planes and algebraic
foldings.

For the technical details, we refer to the two  
papers, \cite{Bordner:1999xs} and \cite{Fuchs:1996zr},
that develop their formalism separately, and we recall merely the
ingredients that enable us to clearly lay down the map between the two
formalisms.\footnote{We will use the notation of  \cite{Fuchs:1996zr} 
throughout this section and refer to that paper and
\cite{Fuchs:1992nq} for our notations.
 The main difference with the body of the paper
is the bar over the simple roots of the simple Lie algebra $g$. The 
index $0$ refers to the zeroth, affine root.}

Indeed, the folding used in \cite{Bordner:1999xs}  corresponds to modding
out an affine algebra by an outer automorphism of the Dynkin diagram
 ($\dot{\omega}$). The action on the Dynkin diagram induces an action 
$\bar{\omega}^{\ast}$ on  the
simple roots, as in 
\cite{Bordner:1999xs}\footnote{Denoted $A$ in that
paper.}. The action on the root system is:
\begin{eqnarray}
\bar{\omega}^{\ast} \bar{\alpha}^{(i)} &=&
 \bar{\alpha}^{(\dot{\omega}i)} \quad \mbox{for} \quad i \neq
 \dot{\omega}^{-1} (0) 
\nonumber \\
\bar{\omega}^{\ast} \bar{\alpha}^{ \dot{\omega}^{-1}(0)}
&=& - \bar{\theta}
\end{eqnarray}
where $\bar{\theta}$ is the highest root of the untwisted affine
lie algebra $g^{(1)}$.
From the framework developed in \cite{Fuchs:1996zr}, it is clear that
 the action of  $\bar{\omega}^{\ast}$ is naturally extended (as an
affine Weyl transformation) to the whole
 weight space. In particular, the action on the Dynkin components
of vectors in the weight space is \cite{Fuchs:1996zr}:
\begin{eqnarray}
(\bar{\omega}^{\ast} \bar{\lambda})^{j} &=&
 (\bar{\lambda})^{{\dot{\omega}}^{-1} (j)} \quad \mbox{for} \quad j \neq
 \dot{\omega} (0) \nonumber \\
(\bar{\omega}^{\ast} \bar{\lambda})^{ \dot{\omega}(0)}
&=& \check{k}_{\lambda}-\sum_{j=1}^r \check{a_{j}} \bar{\lambda}^j.
\label{transfo}
\end{eqnarray}
Now, it is not difficult to see\footnote{Take the inner product of the
transformation rule in \cite{Bordner:1999xs} (3.3) with the
simple roots of $g$.} that these transformation rules 
coincide precisely with the ones guessed 
in \cite{Bordner:1999xs}. In particular
the second transformation rule in (\ref{transfo}) takes into account the  
shift by a fundamental weight that was introduced ad hoc in
 \cite{Bordner:1999xs}. Indeed,
our first gain in this analysis is a systematic derivation of the fact
that in  \cite{Bordner:1999xs} this shift always turned out to 
correspond to the fundamental weight associated to the node $\dot{\omega}(0)$. 
We see now that this has a natural explanation when we realize that the
Dynkin diagram automorphism is actually associated to an affine Weyl 
transformation, as explained in \cite{Fuchs:1996zr}.

We gain a little more when we realize that in \cite{Fuchs:1996zr}
 the orbit algebras for all affine algebras were classified.
Moreover, it was remarked (\cite{Fuchs:1996zr} p. 18)
that the  subalgebra pointwise fixed under the automorphism
has a Cartan matrix
which is the transpose of the Cartan matrix of the orbit 
algebra. Thus we can trivially extend the table in \cite{Fuchs:1996zr} (p. 13)
 to include a column with the
invariant subalgebras. We merely dualize (transpose) the algebras
in their last column. This demonstrates on the one hand that all 
cases in  \cite{Bordner:1999xs} are indeed present in the classification
of \cite{Fuchs:1996zr}, and that we can identify the Cartan torus of
the pointwise fixed algebra with the reduced phase space of \cite{Bordner:1999xs}.
 More importantly, by examining the classification table, we notice
that by folding an affine Dynkin diagram in any other way than the foldings
exhibited in  \cite{Bordner:1999xs}, we will not get a new integrable
system, but we will merely recover the ones we already new. This explains
the systematics behind the folding procedures, and shows that the trial and
error procedure of  \cite{Bordner:1999xs} covered all cases.

It is satisfying to show that the analysis of  \cite{Bordner:1999xs}
 is complete and
to link the folding procedure with the algebra used in fixed
point conformal field theories. This raises the question of whether
there are more applications of this formalism in the realm of
integrable systems. It would be nice if one could make
a precise connection between, say, the application of orbit algebras
in the theory of moduli spaces of flat connections of non-simply
connected gauge groups (as in \cite{Schweigert:1997tg}),
 and  integrable systems on tori.
 A further direction in which to proceed would
be to include current algebras and connections $\bar{A}$ 
with a twist on a two-torus, and examine the integrable systems that
arise from the geometrical setup in algebraic terms.

\section{Conventions}
\label{theta}
\subsection{Theta-functions}
Our conventions are such that
\begin{eqnarray}
\theta_{11}(z+1) &=& - \theta_{11} (z) \\
\theta_{11}(z+ i) &=& e^{- 2 \pi i z} \theta_{11} (z),
\end{eqnarray}
and the $\theta_{11}$ function has zeroes at $n+mi$ with
$n,m \in \mbox{\boldmath $Z$}$.
For simplicity only, we have chosen the periods to be
$\omega_1=1$ and $\omega_2=i$, and we have $\tau=i$.

\end{document}